# Comment on "An arbitrated quantum signature protocol based on the chained CNOT operations encryption"

Yi-Ping Luo[1] and Tzonelih Hwang[*]

*Department of Computer Science and Information Engineering, National Cheng Kung University, No. 1, University Rd., Tainan City, 70101, Taiwan, R.O.C.*

[1] yiping@ismail.csie.ncku.edu.tw

[*] hwangtl@ismail.csie.ncku.edu.tw

## Abstract

In 2015, Li et al. (Quantum Inf Process (2015) 14:2171–2181) proposed an arbitrated quantum signature (AQS) scheme based on the chained controlled-NOT operations encryption. However, this paper points out that in their scheme an attacker can forge a signature without being detected. Therefore, Li et al.'s AQS scheme cannot satisfy the unforgeability and non-repudiation property.

[*] corresponding author

# 1  Introduction

In 2001, Gottesman and Chuang [1] firstly brought out the idea of designing an arbitrated quantum signature (AQS) scheme based on fundamental principles of quantum physics. In their AQS scheme, both the authentication of identities and the integrity of the classical messages or quantum states can be guaranteed over insecure quantum channels. After that, various AQS schemes have been proposed [2-20].

In 2015, Li et al. [21] proposed an AQS scheme based on the chained controlled-NOT (CNOT) operations encryption, which makes each qubit of quantum signature relevant to each other. However, in this paper, we show that, an attacker can forge signer's signature without being detected in Li et al.'s AQS scheme. Therefore, the requirements of **unforgeability** (i.e., neither the signature receiver nor an attacker can forge a signature or change the content of a signature) and **non-repudiation** (i.e., after signing a valid signature, a signer should not be able to deny that) cannot be satisfied in Li et al.'s AQS scheme.

The rest of this paper is organized as follows. Section 2 reviews Li et al.'s AQS scheme. Section 3 describes the forgery attack in Li et al.'s scheme. Section 4 summarizes the result.

# 2  Review of Li et al.'s AQS Scheme

In this section, at first we describe the technique of the chained CNOT operations encryption (in Section 2.1), which is used in Li et al.'s AQS scheme. Hereafter, a brief overview of Li et al.'s AQS scheme is given in Section 2.2.

## 2.1  The Chained CNOT Operations Encryption

Suppose the secret key $K=(k_1,k_2,...,k_n)$ is a permutation of $(1,2,...,n)$, which is shared between the sender and the receiver. The quantum message $|P\rangle$ can be encrypted to $|C\rangle = E_K(|P\rangle)$ by using the chained operations encryption as follows, where $|P\rangle = \otimes_{i=1}^{n}|p_i\rangle$, $|p_i\rangle = \alpha_i|0\rangle + \beta_i|1\rangle$, $\alpha_i, \beta_i \in \mathbf{C}$, $|\alpha_i|^2 + |\beta_i|^2 = 1$, and $1 \leq i \leq n$.

$$|C\rangle = E_K(|P\rangle) = \mathrm{CNOT}(p_n, p_{k_n})\mathrm{CNOT}(p_{n-1}, p_{k_{n-1}})...\mathrm{CNOT}(p_1, p_{k_1})|P\rangle$$

Here, $\mathrm{CNOT}(p_i, p_{k_i})$ denotes as the CNOT operation which takes $p_i$ as a controlled qubit and $p_{k_i}$ as a target qubit. If $k_i = i$, then $\mathrm{CNOT}(p_i, p_i)$ performs the identity operation $I$, where $\mathrm{CNOT} = \begin{bmatrix} 1 & 0 & 0 & 0 \\ 0 & 1 & 0 & 0 \\ 0 & 0 & 0 & 1 \\ 0 & 0 & 1 & 0 \end{bmatrix}$ and $I = \begin{bmatrix} 1 & 0 & 0 & 0 \\ 0 & 1 & 0 & 0 \\ 0 & 0 & 1 & 0 \\ 0 & 0 & 0 & 1 \end{bmatrix}$.

The corresponding decryption method is shown as follows.

$$|P\rangle = D_K(|C\rangle) = \mathrm{CNOT}(p_1, p_{k_1})\mathrm{CNOT}(p_2, p_{k_2})...\mathrm{CNOT}(p_n, p_{k_n})|C\rangle$$

## 2.2 A Brief Review of Li et al.'s AQS scheme

In Li et al.'s AQS scheme, three participants are involved: the signer Alice, the receiver Bob, and the trusted arbitrator Trent. In their scheme, Alice wants to sign the quantum message $|P\rangle$ and transmits it to the signature receiver, Bob, where $|P\rangle = \otimes_{i=1}^{n}|p_i\rangle$, $|p_i\rangle = \alpha_i|0\rangle + \beta_i|1\rangle$, $\alpha_i, \beta_i \in \mathbf{C}$, $|\alpha_i|^2 + |\beta_i|^2 = 1$, and $1 \leq i \leq n$. Subsequently, Bob can verify Alice's signature with the help of Trent. If $|P\rangle$ is composed of known quantum states, arbitrary copies of $|P\rangle$ can be produced. If $|P\rangle$ is composed of unknown quantum states, then at least three copies of $|P\rangle$ should be prepared, i.e. $|P\rangle_1$, $|P\rangle_2$, and $|P\rangle_3$, where $|P\rangle_1 = |P\rangle_2 = |P\rangle_3$. Li et al.'s AQS scheme is composed of three phases: the initializing phase, the signing phase, and the verifying phase.

**Initializing phase**

**Step I1.** Trent shares the secret keys $K_A = \left(k_A^1, k_A^2, ..., k_A^n\right)$ and $K_B = \left(k_B^1, k_B^2, ..., k_B^{2n+1}\right)$ with Alice and Bob, respectively, through the unconditionally secure quantum key distribution protocols, where $\left(k_A^1, k_A^2, ..., k_A^n\right)$ is a permutation of $(1, 2, ..., n)$ and $\left(k_B^1, k_B^2, ..., k_B^{2n+1}\right)$ is a permutation of $(1, 2, ..., 2n+1)$. Besides, Bob shares a secret random bit string $r_B$ with Trent.

**Step I2.** Alice generates $n$ Bell states $|\varphi_1\rangle \otimes |\varphi_2\rangle \otimes ... \otimes |\varphi_n\rangle$, $|\varphi_i\rangle = \frac{1}{\sqrt{2}}\left(|00\rangle_{AB} + |11\rangle_{AB}\right)$, $i = 1, 2, ..., n$, where the particles $A$ and $B$ indicated their owners Alice and Bob, respectively. Then, Alice sends the particle $B$ of each Bell state to Bob through an authenticated method.

**Signing phase**

**Step S1.** Alice randomly chooses a $K_R = \left(k_R^1, k_R^2, ..., k_R^n\right)$, where $\left(k_R^1, k_R^2, ..., k_R^n\right)$ is a permutation of $(1, 2, ..., n)$. Subsequently, Alice transforms $|P\rangle_i$ into $|P_{enc}\rangle_i = E_{K_R}\left(|P\rangle_i\right)$ based on $K_R$ by using the chained CNOT operations encryption, where $1 \leq i \leq 3$.

**Step S2.** Alice generates the signature $|S_A\rangle = E_{K_A}\left(|P_{enc}\rangle_1\right)$ based on $K_A$.

**Step S3.** Alice performs Bell measurement on each $|p_{enc,i}\rangle_2$ of $|P_{enc}\rangle_2$ $(i = 1, 2, ..., n)$ with each particle $A$ of the Bell states to obtain $|M_A\rangle = (|M_A^1\rangle, |M_A^2\rangle, ..., |M_A^n\rangle)$, where $|M_A^i\rangle \in \{|\phi^+\rangle, |\phi^-\rangle, |\psi^+\rangle, |\psi^-\rangle\}$, $|\phi^\pm\rangle = \frac{1}{\sqrt{2}}(|00\rangle \pm |11\rangle)$, $|\psi^\pm\rangle = \frac{1}{\sqrt{2}}(|01\rangle \pm |10\rangle)$, $1 \leq i \leq n$.

**Step S4.** Alice sends $|S\rangle = |P_{enc}\rangle_3 \otimes |S_A\rangle \otimes |M_A\rangle$ to Bob.

**Verifying phase**

**Step V1.** Upon receiving the quantum sequences, Bob performs the unitary operations on $(|P_{enc}\rangle_3 \otimes |S_A\rangle)$ based on $r_B$ to obtain $(|P_{enc}\rangle_3 \otimes |S_A\rangle)' = \otimes_{i=1}^{2n} \sigma_x^{r_B^i} (|P_{enc}\rangle_3 \otimes |S_A\rangle)$. Subsequently, Bob calculates $|Y_B\rangle = E_{K_B}\left((|P_{enc}\rangle_3 \otimes |S_A\rangle)'\right)$ based on $K_B$ and then sends $|Y_B\rangle$ to Trent.

**Step V2.** Trent decrypts $|Y_B\rangle$ based on $K_B$ and obtains $(|P_{enc}\rangle_3 \otimes |S_A\rangle)'$. After that, Trent decrypts $(|P_{enc}\rangle_3 \otimes |S_A\rangle)'$ by using $r_B$ and gets $|P_{enc}\rangle_3 \otimes |S_A\rangle$. Trent transforms $|P_{enc}\rangle_3$ into $|S_T\rangle = E_{K_A}(|P_{enc}\rangle_3)$ based on $K_A$, and then compares $|S_A\rangle$ with $|S_T\rangle$ by using quantum fingerprinting [22]. If $|S_A\rangle = |S_T\rangle$, he/she sets the verification parameter $|V\rangle = |1\rangle$; otherwise, he/she sets $|V\rangle = |0\rangle$.

**Step V3.** Trent recovers $|S_T\rangle$ to obtain $|P_{enc}\rangle_3$, and then sends $|Y_T\rangle = E_{K_B}(|P_{enc}\rangle_3 \otimes |S_A\rangle \otimes |V\rangle)$ to Bob.

**Step V4.** Bob decrypts $|Y_T\rangle$ to obtain $|P_{enc}\rangle_3, |S_A\rangle, |V\rangle$, then he measures $|V\rangle$. If

$|V\rangle = |0\rangle$, he rejects the signature; otherwise, he continues to the next step.

**Step V5.** Based on the technique of quantum teleportation, Bob recovers his particle $B$ into $|P_{enc}\rangle_B$ by performing the corresponding unitary operations which are indicated in $|M_A\rangle$. After that, Bob compares $|P_{enc}\rangle_B$ with $|P_{enc}\rangle_3$. If $|P_{enc}\rangle_B = |P_{enc}\rangle_3$, then Bob informs Alice to publish $K_R$; otherwise, Bob rejects the signature.

**Step V6.** Alice publishes $K_R$.

**Step V7.** Bob decrypts $|P_{enc}\rangle_3$ with $K_R$ and then obtains $|P\rangle_3$. Hence, Bob holds $(|S_A\rangle, K_R)$ as Alice's signature for $|P\rangle_3$ and Trent holds $K_A$ and $K_R$ to judge a possible dispute in the future between Alice and Bob.

## 3 The Forgery Attack

In this section, we point out that a malicious attacker, Eve, is able to forge Alice's signature in Li et al.'s AQS scheme as follows.

Eve generates a fake quantum signature pair $(|S_E\rangle, K_R)$ as Alice's signature for the quantum message $|P_E\rangle$, where $|S_E\rangle = |P_E\rangle = \otimes_{i=1}^{n}|0\rangle$ and $K_R$ is a random permutation which is published in Step V6. Subsequently, Eve provides $(|S_E\rangle, K_R)$ and $|P_E\rangle$ as a signature of $|P_E\rangle$ from Alice to Trent. In order to verify the signature, Trent checks the pair of $(|S_E\rangle, K_R)$ and $|P_E\rangle$. That is, Trent calculates $|S_T\rangle = E_{K_{AT}} E_{K_R} |P_E\rangle$ and compare $|S_T\rangle$ with $|S_E\rangle$, where $E$ denotes the chained CNOT operations encryption and $K_A$ denotes the shared secret key between Alice

and Trent. Since, the CNOT operation does not affect the states of $\otimes_{i=1}^{n}|0\rangle$, therefore, $|S_T\rangle$ is equal to $|S_E\rangle$. That denotes the signature pair $(|S_E\rangle, K_R)$ and $|P_E\rangle$ did come from Alice.

Therefore, Li et al.'s AQS scheme cannot satisfy the requirement of unforgeability. As a signature is possible to be forged, the signer could also deny a valid signature for his/her own benefit. As a result, the non-repudiation property in a signature scheme is not satisfied.

## 4  Conclusions

In this article, we have pointed out a security loophole in Li et al.'s AQS scheme by using the forgery attack. It would be a challenge to design a secure arbitrated quantum signature scheme.

## Acknowledgment

We would like to thank the Ministry of Science and Technology of Republic of China for financial support of this research under Contract No. MOST 104-2221-E-006-102 -.